\documentclass[aps,prl,twocolumn,showpacs,superscriptaddress,floatfix]{revtex4-1}
\usepackage{amsmath,amssymb}
\usepackage{graphicx}
\usepackage{epsfig}
\usepackage{color}
\usepackage{rotating}
\usepackage{bm}
\usepackage{setspace}
\usepackage{hyperref}

\begin{document}
\title{Nodeless pairing in superconducting copper-oxide monolayer films on Bi$_2$Sr$_2$CaCu$_2$O$_{8+\delta}$}
\author{Yong Zhong}
\thanks{Authors equally contributed to this work.}
\author{Yang Wang}
\thanks{Authors equally contributed to this work.}
\author{Sha Han}
\author{Yan-Feng Lv}
\author{Wen-Lin Wang}
\author{Ding Zhang}
\author{Hao Ding}
\author{Yi-Min Zhang}
\affiliation{State Key Laboratory of Low-Dimensional Quantum Physics, Department of Physics, Tsinghua University, Beijing 100084, China}
\author{Lili Wang}
\author{Ke He}
\affiliation{State Key Laboratory of Low-Dimensional Quantum Physics, Department of Physics, Tsinghua University, Beijing 100084, China}
\affiliation{Collaborative Innovation Center of Quantum Matter, Beijing 100084, China}
\author{Ruidan Zhong}
\author{John Schneeloch}
\author{Gen-Da Gu}
\affiliation{Condensed Matter Physics and Materials Science Department, Brookhaven National Laboratory, Upton, New York 11973, USA}
\affiliation{Department of Physics Astronomy, Stony Brook University, Stony Brook, NY 11794, USA}
\author{Can-Li Song}
\email[]{clsong07@mail.tsinghua.edu.cn}
\author{Xu-Cun Ma}
\email[]{xucunma@mail.tsinghua.edu.cn}
\author{Qi-Kun Xue}
\email[]{qkxue@mail.tsinghua.edu.cn}
\affiliation{State Key Laboratory of Low-Dimensional Quantum Physics, Department of Physics, Tsinghua University, Beijing 100084, China}
\affiliation{Collaborative Innovation Center of Quantum Matter, Beijing 100084, China}
\date{\today}

\begin{abstract}

\end{abstract}

\maketitle

\textbf{Abstract} The pairing mechanism of high-temperature superconductivity in cuprates remains the biggest unresolved mystery in condensed matter physics. To solve the problem, one of the most effective approaches is to investigate directly the superconducting CuO$_2$ layers. Here, by growing CuO$_2$ monolayer films on Bi$_2$Sr$_2$CaCu$_2$O$_{8+\delta}$ substrates, we identify two distinct and spatially separated energy gaps centered at the Fermi energy, a smaller U-like gap and a larger V-like gap on the films, and study their interactions with alien atoms by low-temperature scanning tunneling microscopy. The newly discovered U-like gap exhibits strong phase coherence and is immune to scattering by K, Cs and Ag atoms, suggesting its nature as a nodeless superconducting gap in the CuO$_2$ layers, whereas the V-like gap agrees with the well-known pseudogap state in the underdoped regime. Our results support an s-wave superconductivity in Bi$_2$Sr$_2$CaCu$_2$O$_{8+\delta}$, which, we propose, originates from the modulation-doping resultant two-dimensional hole liquid confined in the CuO$_2$ layers.
\\
\newline
\newline
\textbf{1 Introduction}
\newline

The discovery of high temperature ($T_\textrm{c}$) superconductivity (HTS) in cuprates \cite{bednorz1986possible} has triggered tremendous efforts to elucidate as to why they superconduct at high \emph{T}$_c$. However, this enigma remains unresolved, apparently owing to the layered structure in which the superconducting CuO$_2$ layers are sandwiched between non-superconducting charge reservoir layers, for example, BiO/SrO in the case of Bi$_2$Sr$_2$CaCu$_2$O$_{8+\delta}$(Bi-2212). Upon chemical doping in the reservoir layers, rather than direct chemical doping in the CuO$_2$ layers, the resulting carriers transfer into the CuO$_2$ layers and boost superconductivity therein, resembling the HTS in single-unit-cell FeSe films on SrTiO$_3$ \cite{qing2012interface}. By layer-by-layer removal with an Ar$^+$ bombardment technique, we could reveal respective quasiparticle (QP) excitations of the constituent planes of Bi-2212 \cite{Lv2015mapping} and Bi-2201 \cite{Lv2016electronic}. However, the CuO$_2$ surface obtained is too small to systematically investigate its superconducting property.

Here we grow ultrathin CuO$_2$ films on the BiO surfaces of the cleaved Bi-2212 crystals by a state-of-the-art ozone molecular beam epitaxy (MBE) technique. This approach allows direct measurements of the electronic structure of the CuO$_2$ layers by \textit{in situ} scanning tunneling microscopy (STM). We demonstrate that the superconducting gap in the CuO$_2$ layers is nodeless, contradictory to the nodal \emph{d}-wave pairing scenario that is often thought to be the most important result in the thirty-year study of the HTS mechanism of cuprates.
\\
\newline
\newline
\textbf{2 Experimental}
\newline

Our experiments were conducted in an ultrahigh vacuum low temperature STM system equipped with ozone-assisted MBE chamber (Unisoku), with a base pressure of 1$\times$10$^{-10}$ Torr. The copper-oxide films were prepared by evaporating high-purity Cu (99.9999\%) sources from a standard Knudsen cell under ozone flux beam of 1.0 $\sim$ 5.0$\times$10$^{-5}$ Torr. The flux beam of ozone from a home-built ozone gas delivery system was injected into the MBE using a 1/2 inch stainless tube (Swagelok), $\sim$ 50 mm distant from the sample. The K, Cs and Ag atoms were respectively evaporated from alkali-metal dispensers (SAES Getters) and home-made Ta boat, with the samples kept at approximately 50 K. Polycrystalline PtIr tips were cleaned by $e$-beam heating in UHV and calibrated on the MBE-grown Ag/Si(111) films before STM measurements. All STM images were acquired at 4.2 K in constant-current mode, and the differential tunneling \emph{dI/dV} spectra were measured by using a standard lock-in technique with a small bias modulation of 1 mV at 966 Hz, unless otherwise specified.
\\
\newline
\newline
\textbf{3 Result and discussion}
\newline

\begin{figure*}[tbh]
\includegraphics[width=0.975\textwidth]{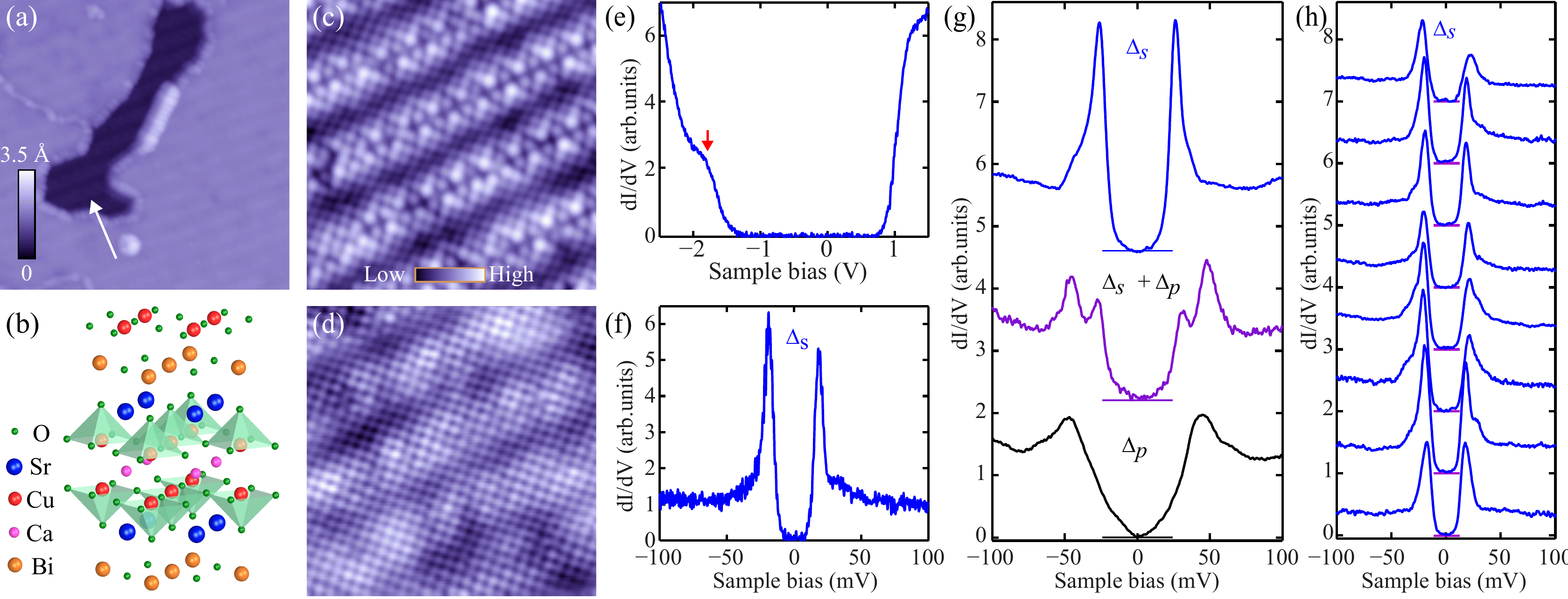}
\caption{STM characterization of CuO$_2$ monolayer films on Bi-2212. \textbf{a, b} STM topography (\emph{V} = 1.5 V, \emph{I} = 20 pA, 40 nm $\times$ 40 nm) and schematic view of monolayer CuO$_2$ films on the Bi-2212 substrate. The arrow indicates the uncovered BiO surface. \textbf{c} STM topography (\emph{V} = 0.2 V, \emph{I} = 50 pA, 10 nm $\times$ 10 nm) on the BiO surface. \textbf{d} STM topography (\emph{V} = 0.5 V, \emph{I} = 15 pA, 10 nm $\times$ 10 nm) of the CuO$_2$ films showing a square lattice with Cu-Cu spacing of 3.8 \AA. \textbf{e} Large-energy-scale \emph{dI/dV} spectrum (setpoint: \emph{V} = 1.5 V, \emph{I} = 100 pA) on CuO$_2$. The bias-modulation amplitude was set to 15 mV$_\textrm{{rms}}$. \textbf{f, g} \emph{dI/dV} spectra of the CuO$_2$ films, revealing two distinct energy scale gaps ($\Delta_s$ and $\Delta_p$) in the low-energy QP excitations. The U-shaped spectra exhibit vanishing and particle-hole-symmetric DOS over a finite energy range near $E_F$. The spectra in (\textbf{g}) were acquired in the same sample and vertically offset for clarity, with their zero-conductance positions marked by the correspondingly colored horizontal lines throughout. \textbf{h} Line-cut \emph{dI/dV} spectra showing the spatial robustness of the U-shaped superconducting gap in the CuO$_2$ films. The eight \emph{dI/dV} spectra along 6-nm trajectories on CuO$_2$ all show vanishing and particle-hole-symmetric DOS over a finite energy range near $E_F$. The purple horizontal bars mark the zero-conductance positions. Setpoint: \textbf{f, h} \emph{V} = 0.2 V, \emph{I} = 100 pA; \textbf{g} \emph{V} = 0.1 V, \emph{I} = 300 pA.}
\end{figure*}

MBE growth of copper-oxide films on Bi-2212 with \emph{T}$_c$ = 91 K (Fig.\ \textbf{1}a-d) proceeds in layer-by-layer mode, and epitaxial crystalline films down to a single monolayer (ML) could be prepared. The STM topographic images (Fig.\ \textbf{1}a, d) reveal atomically flat and nearly defect-free surfaces on the cleaved Bi-2212 crystal. The characteristic \emph{b}-axis supermodulation (Fig.\ \textbf{1}c) is clearly seen in the uncovered BiO regions in Fig.\ \textbf{1}a. The atomically resolved image in Fig.\ \textbf{1}d reveals the nearest-neighboring Cu ions in the monolayer films to be 3.8 \AA \space apart. The square structure and the in-plane lattice constant are consistent with those of the CuO$_2$ plane in bulk Bi-2212. For rocksalt-structured tetragonal CuO (T-CuO) \cite{Siemons2009tetragonal, Moser2014angle, Adolphs2016non}, a $\sqrt{2}$$\times$$\sqrt{2}$ surface reconstruction might blur half Cu ions and lead to the same lattice constant. However, the identification of the films as CuO$_2$ is sustained by a large separation of 2.35 eV between the Fermi level ($E_F$) and the valence band of T-CuO \cite{Moser2014angle}, while the corresponding value here is only 1.40 eV (Fig.\ \textbf{1}e).

In scanning tunneling spectroscopy (STS), the differential conductance \emph{dI/dV} measures the local density of states (DOS) as a function of bias voltage (\textit{V}). Our \emph{dI/dV}  measurements reveal several distinctive features of the CuO$_2$ films from the cleaved BiO surfaces that have been extensively studied by STM/STS \cite{fischer2007scanning}. First, the overall electronic spectrum of the CuO$_2$ films is characterized by a Mott-Hubbard-like gap of 2.21 eV without prominent states in it (Fig.\ \textbf{1}e). The result is expected since the ground state of CuO$_2$ is a Mott insulator as previously observed by STM in Ca$_2$CuO$_2$Cl$_2$ \cite{ye2013visualizing}. The pronounced DOS at 1.1 eV can be assigned to the upper Hubbard band, while the shoulders around -1.82 eV (see the arrow in Fig.\ \textbf{1}e) the charge transfer band (CTB) \cite{PhysRevLett.55.418} or Zhang-Rice-Singlet \cite{PhysRevB.37.3759} because of strong on-site Coulomb repulsion. The CTB gap magnitude of 2.21 eV agrees with 2.2 eV as measured in Ca$_2$CuO$_2$Cl$_2$ [9] and 2.0 eV for La$_{2-x}$Sr$_x$Cu$_2$O$_4$ \cite{RevModPhys.70.897}.

\begin{figure}[b]
\includegraphics[width=0.5\textwidth]{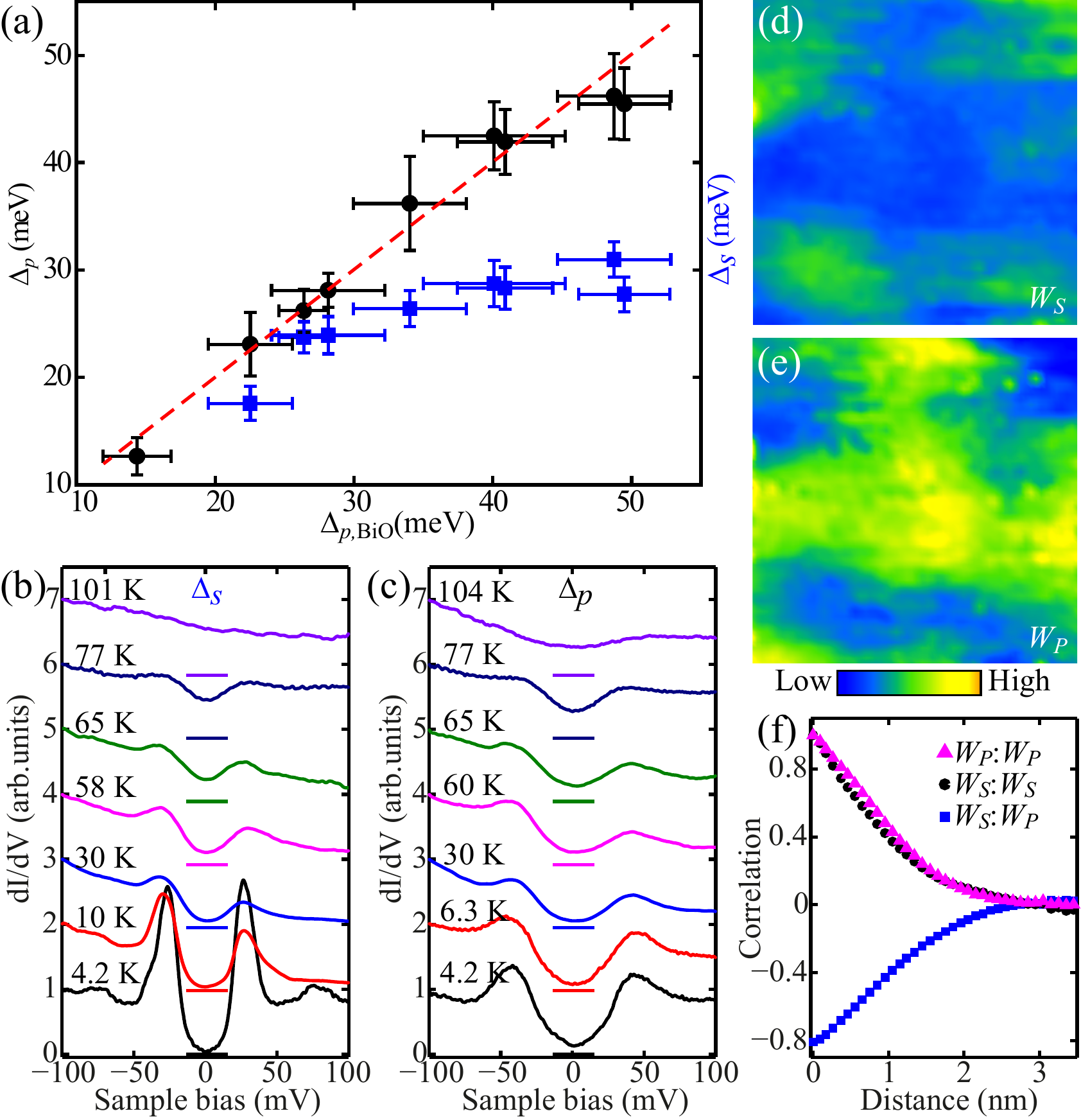}
\caption{\noindent Low-energy QP excitations at two energy scales. \textbf{a} Dependence of $\Delta_s$ (blue squares) and $\Delta_p$ (black circles) on $\Delta _{p, \textrm{BiO}}$. The red diagonal dashes show the curve for $\Delta_p$ = $\Delta _{p, \textrm{BiO}}$. The error bars refer to the standard deviation of statistical gap values from 50 \emph{dI/dV} spectra. \textbf{b, c} Temperature dependence of the superconducting gap $\Delta_s$ and PG $\Delta_p$. Distinct from a gradual DOS filling for $\Delta_p$, the DOS near $E_F$ for $\Delta_s$ shows a more dramatic increase with temperature. Setpoint: \emph{V} = 0.1 V, \emph{I} = 100 pA. \textbf{d, e} Local DOS images (6 nm $\times$ 6 nm) of \emph{dI/dV} spectra measured at (\textbf{d}) \emph{V} = -25 mV and (e) \emph{V} = -45 mV, characterizing the spatial distribution of $W_{s(p)}$ for the pairing gap $\Delta_s$ and pseudogap $\Delta_p$, respectively. Setpoint: \emph{V} = 50 mV, \emph{I} = 100 pA. \textbf{f} Azimuthally averaged autocorrelation of $W_{s(p)}$ (black circles and magenta triangles) and cross correlation between them (blue squares).}
\end{figure}

Second, and the most surprisingly, in some regions the low-energy QP excitations of the CuO$_2$ films disclose a well-defined U-gap (Fig.\ \textbf{1}f) ($\Delta_s$). Its two pronounced conductance peaks at gap edges ($\pm$ 18.6 meV) and the vanishing DOS in between, suggest that $\Delta_s$ is an \emph{s}-wave-like gap. To avoid possible artificial effects in our measurements, we collected more than 3000 spectra on different locations of the CuO$_2$ films from more than forty samples. The persistence of the U-gap in these regions is illustrated in Fig.\ \textbf{1}h: despite changes in the coherence peak intensity and the gap size ranging from 16 meV to 22 meV, the essential U shape can clearly be seen. The observation invalidates the general assumption that the V-shaped STS spectra on the cleaved BiO surfaces are due to tunneling from the underlying CuO$_2$ planes in Bi-2212 \cite{fischer2007scanning}. Other regions are characterized by a larger V-shaped gap ($\Delta_p$) (the black curve in Fig.\ \textbf{1}g). At boundaries between the two types of regions, a double-gap feature with mixed U- and V-gaps (violet curve in Fig.\ \textbf{1}g) is present and expected. Within our instrumental resolution, we find no difference in the detailed features of these V-gaps and the pseudogap (PG) from the BiO surfaces in Fig.\ \textbf{1}c \cite{fischer2007scanning}. The two-gap feature, together with the systematic spatial evolution of \emph{dI/dV} spectra observed throughout, suggests two types of QP gaps at different energy scales in the CuO$_2$ films on Bi-2212, although their spectral weights (\emph{W}) are spatially varied. Such spatial inhomogeneity might originate from different amount of charge transfer from the substrate to CuO$_2$, e.g.\ the U-gap regions have more hole carriers as compared to the V-gap regions. This charge transfer seems not induce apparent large-energy-scale states in the CTB gap (Fig.\ \textbf{1}e). The result is very significant and implies that external doping of the CuO$_2$ Mott insulator does not alter its fundamental electronic structure, an issue fiercely debated but at the heart of HTS physics. It actually forms a starting point of our new model for HTS, as will be discussed later.

What might be the origin of the two-energy-scale QP gaps in the CuO$_2$ films? To address this issue, in Fig.\ 2a, we summarize our measurements of the averaged gap magnitudes, $\Delta_s$ and $\Delta_p$, as a function of $\Delta_{p, \textrm{BiO}}$ simultaneously acquired on the exposed BiO surfaces. Here $\Delta_{p, \textrm{BiO}}$ and thus the hole carriers are delicately controlled by annealing the samples under ozone flux or UHV \cite{Lv2015mapping, Lv2016electronic}. The larger energy gap, $\Delta_p$, is found to scale linearly with $\Delta_{p, \textrm{BiO}}$ and bears striking quantitative correspondence in the magnitude (red dashes) and spatial variation (marked by the equal statistical uncertainty) to $\Delta_{p, \textrm{BiO}}$, further stressing its nature as the gap in Bi-2212 substrates \cite{hashimoto2014energy}. Therefore, the CuO$_2$ films in these regions remain at their insulating state and thus transparent to the STM tip. On the other hand, the magnitude of the smaller but relatively homogeneous gap $\Delta_s$ (documented by the small statistical uncertainty in Fig.\ \textbf{2}a) exhibits a different behavior versus $\Delta_{p, \textrm{BiO}}$ and saturates at $\sim$ 30 meV. This hints at a different origin of $\Delta_s$. As a superconducting gap in the CuO$_2$ layers, it was occasionally seen in the form of broad \lq kinks\rq \space in underdoped Bi-2212 \cite{pushp2009extending, alldredge2008evolution}. The above results including those from Fig.\ \textbf{1}f, g, i, consistently suggest that the U-gap corresponds to a superconductivity state prompted by the charge transfer mentioned above. Because of asymmetric doping and a peculiar bonding configuration of CuO$_2$ on the inert BiO surface, we speculate that the total charge transfer is not sufficient to make the whole films superconducting, leading to the formation of two different regions. The situation may be improved by growing CuO$_2$ on the SrO surface of Bi-2212 \cite{Lv2015mapping, Lv2016electronic}.

To further confirm that the U-gap regions are superconducting, we investigate the temperature evolution of the \emph{dI/dV} spectra (see Fig.\ \textbf{2}b, c). The U-gap $\Delta_s$ exhibits the characteristics of a conventional superconducting gap: with increasing temperature, the coherence peaks begin to fade away abruptly and then gradually (Fig.\ \textbf{2}b). At 101 K, the gap vanishes completely. Meanwhile, the DOS near $E_\emph{F}$ shows a dramatic increase for $\Delta_s$, particularly around 101 K. The results demonstrate that $\Delta_s$ is indeed a superconducting gap, and that the CuO$_2$ monolayers on BiO surface are superconducting with \emph{s}-wave-like pairing wavefunction. As for $\Delta_p$, it is still visible at 104 K and exhibits the same behavior as the PG observed on the cleaved Bi-2212 surfaces \cite{fischer2007scanning}.

\begin{figure*}[tbh]
\includegraphics[width=0.99\textwidth]{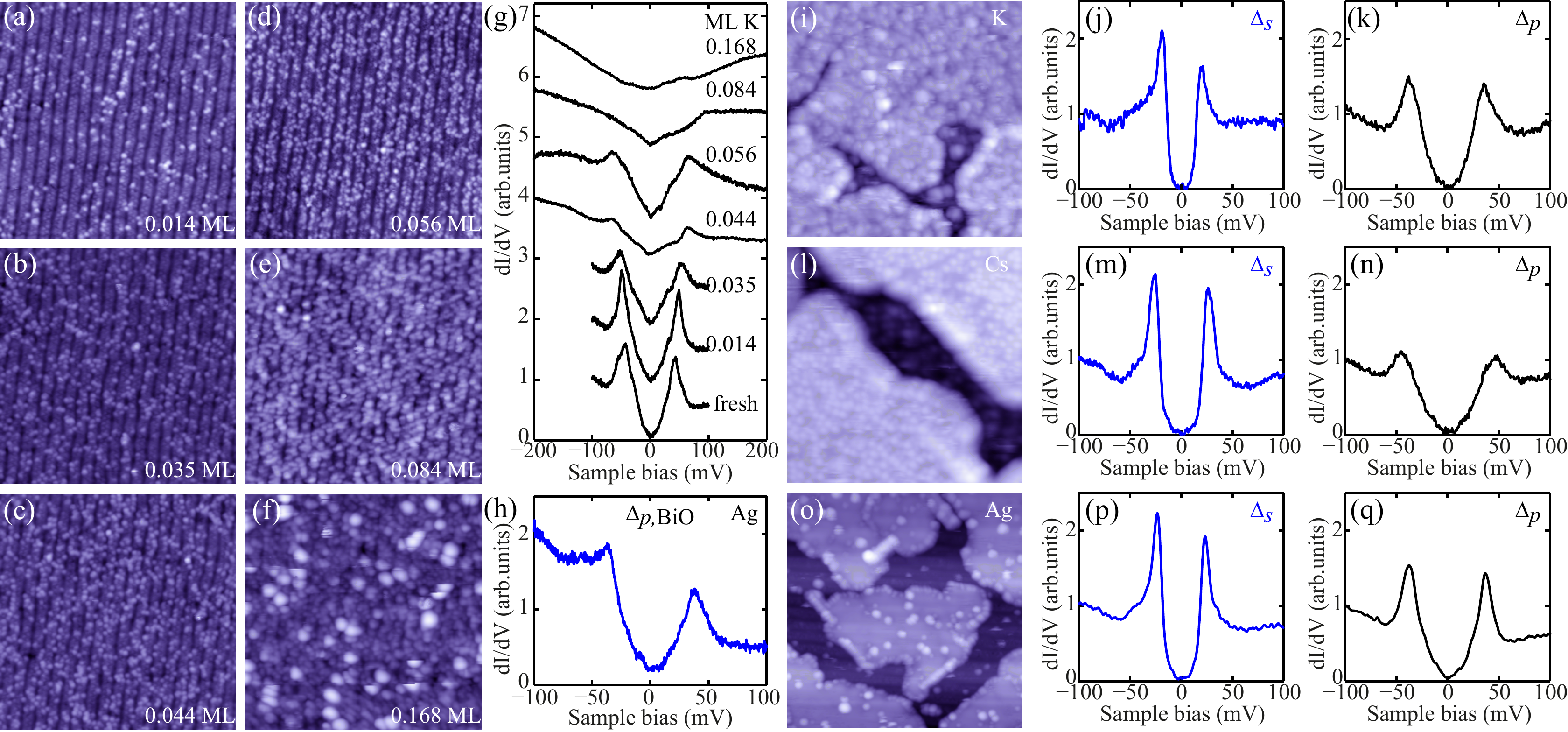}
\caption{\noindent Immunity of QP energy gaps to non-magnetic impurities. \textbf{a-f} STM topography (\emph{V} = 0.4 V, \emph{I} = 20 pA, 50 nm $\times$ 50 nm) of the cleaved Bi-2212 surface adsorbed with K at different dosages, as indicated. Here 1 ML is defined as the areal density of Bi at the topmost BiO layer ($\sim$ 6.9 $\times$ 10$^{14}$/cm$^2$). \textbf{g} K coverage-dependent \emph{dI/dV} spectra on the cleaved Bi-2212 surface at 4.2 K. Setpoint: \emph{I} = 200 pA and \emph{V} = 0.1 V (the three curves from the bottom) or 0.2 V (the upper four curves). \textbf{h} Typical \emph{dI/dV} spectrum (setpoint: \emph{V} = 0.2 V, \emph{I} = 200 pA) measured on Ag adatoms on Bi-2212, revealing no in-gap states within the pseudogap. \textbf{i-k} STM topography (\emph{V} = 1.2 V, \emph{I} = 10 pA, 50 nm $\times$ 50 nm) and \emph{dI/dV} spectra (setpoint: \emph{V} = 0.2 V, \emph{I} = 200 pA) of $\sim$ 0.84 ML K-dosed CuO$_2$ monolayer films. \textbf{l-n} STM topography (\emph{V} = 1.5 V, \emph{I} = 10 pA, 30 nm $\times$ 30 nm) and \emph{dI/dV} spectra (setpoint: \emph{V} = 0.2 V, \emph{I} = 200 pA) of $\sim$ 0.03 ML Cs-dosed CuO$_2$ monolayer films. \textbf{o-q} STM topography (\emph{V} = 1.2 V, \emph{I} = 20 pA, 50 nm $\times$ 50 nm) and \emph{dI/dV} spectra (setpoint: \emph{V} = 0.2 V, \emph{I} = 200 pA) of Ag-dosed CuO$_2$ monolayer films. The spectra were measured on K, Cs or Ag adatoms, revealing undisturbed energy gaps against nonmagnetic impurity scattering.}
\end{figure*}

In search for the origin of the nodeless superconductivity observed here, we study the related spectral weight of $\Delta_{s(p)}$, \emph{$W_{s(p)}$}, in the CuO$_2$ films. Such \emph{$W_{s(p)}$} images, represented approximately by the \emph{$dI/dV$} conductance at the energies of $\Delta_s$ $\sim$ 25 meV and $\Delta_p$ $\sim$ 45 meV, are simultaneously mapped and shown in Fig.\ \textbf{2}d, e. The spatial inhomogeneity in \emph{$W_{s(p)}$} and their anticorrelation are discernable. To quantify this, we compute the cross correlation between \emph{$W_s$} and \emph{$W_p$}, as depicted by blue squares in Fig.\ \textbf{2}f. Combined with the \emph{$W_{s(p)}$} autocorrelation (black circles and magenta triangles), we estimate a characteristic correlation length of $\sim$ 2.7 nm over which the correlation goes to zero. The findings are reminiscent of the electronic inhomogeneity with a similar length scale in cuprates \cite{pan2001microscopic, mcelroy2005atomic}, which is most likely correlated with interstitial oxygen dopants \cite{mcelroy2005atomic}. The anticorrelation coefficient of -0.8 between \emph{$W_s$} and \emph{$W_p$} is actually quite large, suggesting that the Bi-2212 substrate plays an important role in the observed superconductivity as a charge reservoir. Evidently, the superconducting gap $\Delta_s$ goes hand in hand with $\Delta _{p, BiO}$ (Fig.\ \textbf{2}a), an indicator of mobile hole carriers available in the Bi-2212 \cite{hufner2008two}. Here the anticorrelated $\Delta_p$ and $\Delta_s$, supporting the proposal that the PG in cuprates is competitive with and thus irrelevant to pairing \cite{Lv2015mapping, Lv2016electronic, kondo2009competition, chakravarty2001hidden}, might be only a phenomenological reflection of the available hole carriers for PG formation and Cooper pairing.

Response of the gaps against nonmagnetic impurities further helps unravel the pairing symmetry. Nonmagnetic impurities have been known to exert little or no effect on electron pairing in a conventional $s$-wave superconductor \cite{yazdani1997probing}, but to induce QP bound states within the pairing gap in an unconventional superconductor with gap nodes \cite{balatsky2006impurity}. To gain insight into $\Delta_p$ and $\Delta_s$, we deposit alkali-metals (K and Cs) and a noble-metal (Ag) on Bi-2212 and CuO$_2$ films (see Fig.\ \textbf{3}). We find that the PG on Bi-2212 broadens gradually in its magnitude with increasing K (Cs) coverage and vanishes at a K dose of $\sim$ 0.168 ML (Fig.\ \textbf{3}g), whereas it changes little on the K, Cs and Ag covered CuO$_2$ surface (Fig.\ \textbf{3}k, n, q). This nicely echoes our argument that the PG is from the substrate: K (Cs) may donate some electrons to the CuO$_2$ surface \cite{kim2014fermi, kim2016observation}, but it is limited to the CuO$_2$ layers (Fig.\ \textbf{3}a-f) and should not disturb the substrate electronic structure much. No bound state is seen atop Ag adatoms on the Bi-2212 surface (Fig.\ \textbf{3}h), implying that the \emph{d}-wave PG is not a pairing gap, which should otherwise induce in-gap bound states \cite{balatsky2006impurity}. Importantly, the long hidden superconducting gap $\Delta_s$ for CuO$_2$ shows striking robustness against scattering by K, Cs (Fig.\ \textbf{3}a, b, d, e), and Ag (Fig.\ \textbf{3}g, h) for a wide range of coverage from 0 to 0.84 ML. The results are in line with the nodeless \emph{s}-wave pairing and against sign-reversal pairing symmetry.

\begin{figure}[t]
\includegraphics[width=0.5\textwidth]{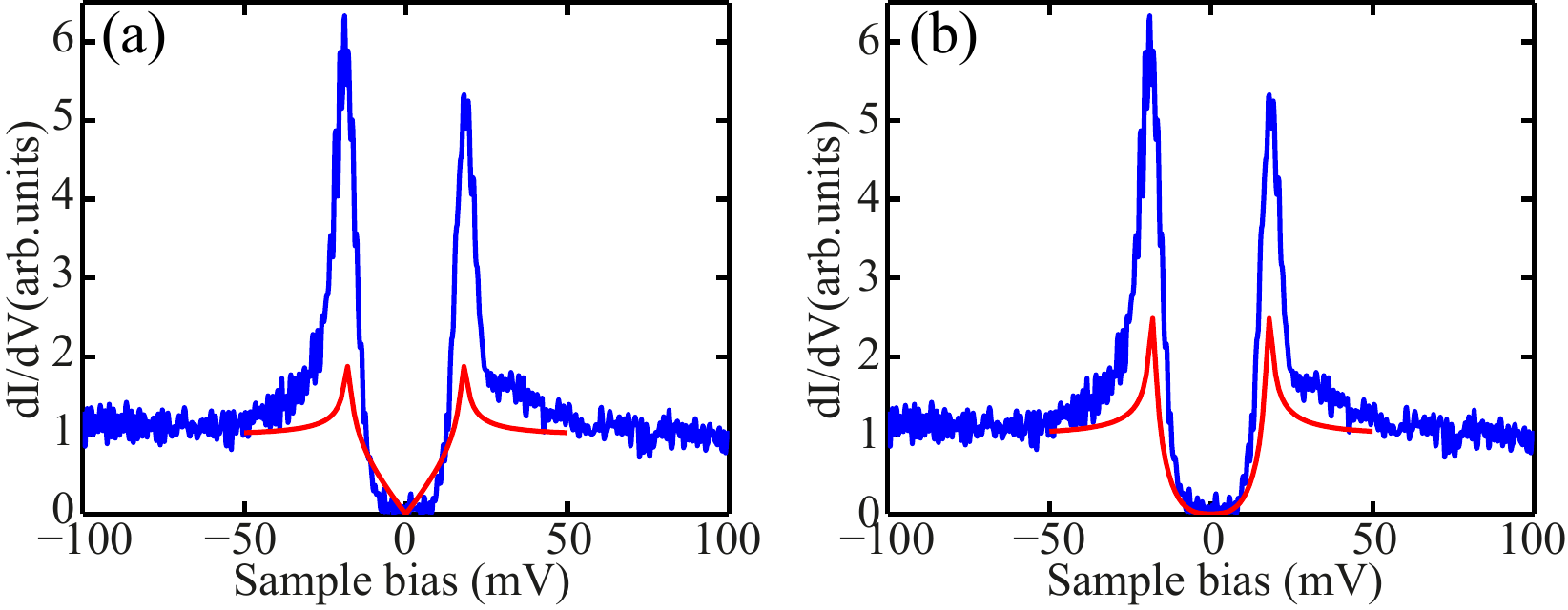}
\caption{Fits (red curve) of the U-like \emph{dI/dV} spectrum of Fig.\ \textbf{1}f (blue curve) to (\textbf{a}) a standard \emph{d}-wave and (\textbf{b}) \emph{T}$_\phi$-filtered \emph{d}-wave. The experimental spectrum shows remarkably strong coherence peaks and cannot be reconciled with a \emph{d}-wave gap even after an inclusion of the anisotropic matrix effects. An instrumental broadening parameter $\Gamma$ = 0.1 meV is used for the fits.}
\end{figure}

\begin{figure}[b]
\includegraphics[width=0.5\textwidth]{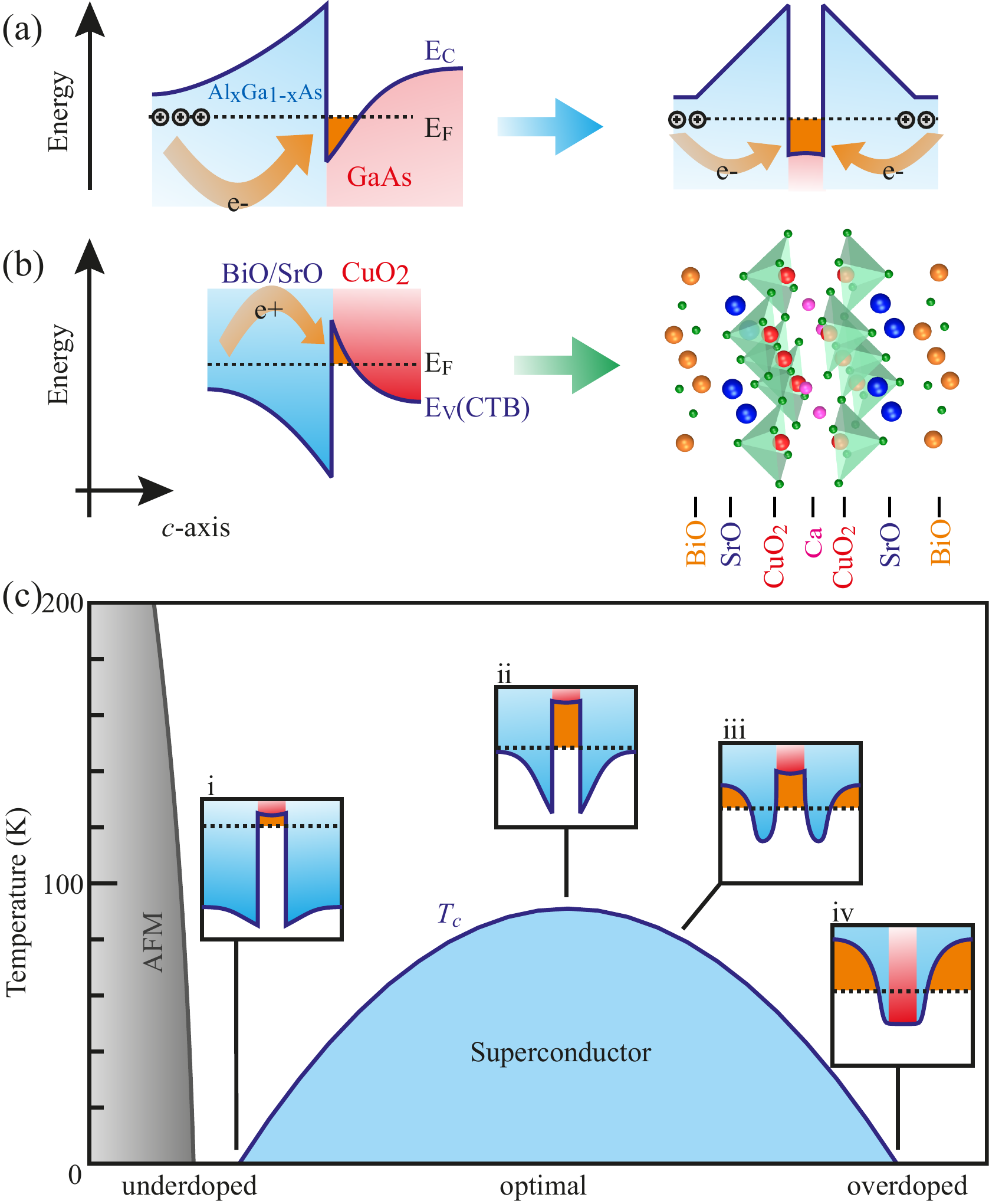}
\caption{Modulation doping scheme and superconductivity in cuprates. \textbf{a, b} Energy-band diagram and formation of quantum wells for Al$_x$Ga$_{1-x}$As/GaAs and SrO(BiO)/CuO$_2$ heterostructures, respectively. Mobile electrons or holes are transferred from the modulation doping layers to either the interface or the quantum wells (indicated by the orange arrows). \textbf{c} Phase diagram of high-\emph{T}$_c$ cuprate superconductors, interpreted as the evolution of a quantum well under modulation doping of holes.}
\end{figure}

One may argue that lifting of gap nodes by disorder might engender a U-shaped gap structure. For hypothesized \emph{d}-wave pairing in cuprates, the positions and topology of nodes are rigorously imposed by the symmetry, and thus the gapless excitations are protected against disorder. In this case, the disorder would have severely suppressed the coherence peaks, contradicting our observations (Fig.\ \textbf{1}f, g, i). One may also argue that the U-gap results from the anisotropic tunneling matrix element \emph{T}$_\phi$ $\propto$ $\cos^2(2\phi)$ ($\phi$ = $\tan^{-1}(k_y/k_x)$ is the azimuthal tunneling angle) effects for CuO$_2$ with tetragonal symmetry \cite{PhysRevB.73.134510}. Such a \emph{T}$_\phi$ reduces the low-energy QP tunneling in the nodal regions of \emph{d}-wave gap and can give rise to a U-like spectrum, while sharpening the coherence peaks. We have attempted the method to fit our data, but no success (Fig.\ \textbf{4}), which suggests that $\Delta_s$ is isotropic in reciprocal space. Further theoretical study is needed to understand how the pairing is intertwined with PG \cite{mishra2014effect} and the resultant U-like excitations in CuO$_2$.

Our results on the nodeless gap of the superconducting CuO$_2$ monolayers, which is evidently triggered by charge transfer from the substrate, have significant implications on the pairing mechanism of Bi-2212. The undoped CuO$_2$ is well known as a Mott insulator due to many-body correlation. It becomes superconducting only when the adjacent BiO/SrO reservoir layers are doped and the resulting carriers are injected into it. Such charge-transfer mechanism has long been employed as modulation-doping in semiconductor high-electron-mobility-transistors such as Al$_x$Ga$_{1-x}$As/GaAs superlattices \cite{dingle1978electron}, as schematically illustrated in Fig.\ \textbf{5}a. It is well-established that charge transfer is initiated only when the $E_F$ of doped n-type Al$_x$Ga$_{1-x}$As is higher than that of undoped GaAs. Upon contact, charge transfer and band bending take place to align their Fermi levels, which, in turn, leads to formation of a quantum well (QW) at the GaAs side of the interface. The transferred carriers form a two-dimensional electron gas (2DEG) in the QW, while Al$_x$Ga$_{1-x}$As is depleted forming a potential barrier. Without band bending and QW formation, GaAs itself is insulating and cannot accommodate electrons. Similarly, the holes cannot be injected into the CuO$_2$ Mott insulator unless a QW forms in the CuO$_2$ layer (Fig.\ \textbf{5}b). As a result, the hole carriers generated in the SrO(BiO) layers \lq\lq fall\rq\rq \space into the CuO$_2$ QW and form a clean 2D hole liquid (2DHL) therein. Formation of 2DHL does not necessarily change the ground electronic state of CuO$_2$, as in the case of Al$_x$Ga$_{1-x}$As/GaAs in the spirit of rigid-band model. This argument is supported by our observations in Fig.\ \textbf{1}e, and also by resonant inelastic X-ray scattering study showing the persistence of anti-ferromagnetic ordering in CuO$_2$ for the whole phase diagram of La$_{2-x}$Sr$_x$Cu$_2$O$_4$ ($x$ = 0 to 0.40) \cite{dean2013persistence}. In this way, regardless of how much the doping is, the CuO$_2$ planes can sustain its chemical and structural integrity so as to assure the superconductivity phase coherence. Under this context, Bi-2212 is nothing special but just a 2DHL superlattice with much higher carrier density. The 2DHL confined in the CuO$_2$ layers should have a simple concentric circular Fermi surface, which may naturally lead to an isotropic pairing and a conventional \emph{s}-wave gap. Indeed, the single-unit-cell FeSe films on SrTiO$_3$ are found to exhibit a simple Fermi surface and conventional \emph{s}-wave superconductivity \cite{qing2012interface, fan2015plain}, which result from the same charge transfer mechanism.

We now show that with this rather simple model, we can understand the complicated phase diagram of cuprates \cite{keimer2015quantum}. As schematically illustrated in Fig.\ \textbf{5}c, at critical doping the superconductivity sets in as the CuO$_2$ QW is filled with sufficient holes to form 2DHL (panel i in Fig.\ \textbf{5}c). Increasing doping concentration pulls down the $E_F$ of the reservoir layers such that the hole density in the QW increases (so does $T_\textrm{c}$). The optimal doping corresponds to the situation where the QW reaches its maximal capacity to accommodate holes (panel ii in Fig.\ \textbf{5}c). Further doping into the reservoir layers results in a parallel conductive channel and the CuO$_2$ QW starts to collapse (panel iii in Fig. 5c), which reduces the effective hole density in the CuO$_2$ layer and thus weakens the superconductivity. In the extremely overdoped regime where the \emph{E}$_F$ touches the CTB of CuO$_2$ (panel iv in Fig.\ \textbf{5}c), the charge transfer no longer exists. As a result, the QW is not filled and the CuO$_2$ layer returns to its initial insulating state, while the reservoir layers become highly conductive. Note that the charge carriers generated in the reservoir layers increase monotonically with chemical doping (Fig.\ \textbf{5}c). It is consistent with the linear reduction of $\Delta_p$ with increasing doping and supports the origin of PG from the reservoir layers.

The strange metal phase around the overdoped regime is regarded as one of the most mysterious puzzles in cuprates \cite{keimer2015quantum}. According to our model, the Bi-2212 in this regime (panel iii in Fig.\ \textbf{5}c) is composed of alternative 2DHL layers and SrO/BiO reservoir layers that are both conducting. Unlike a conventional metal, the carrier density of these layers can be thermally activated, just like doped semiconductors. This fact would complicate the resistivity versus temperature behavior, and the peculiar linearly temperature-dependent resistivity in cuprates results simply from the intertwined effects of a quadratic scattering rate (Fermi liquid behavior) and temperature-dependent carrier density \cite{barivsic2015hidden}. Moreover, other normal state properties of cuprates in the whole phase diagram can be understood easily: they should also be dominated by complicated behaviors of the heavily doped charge reservoir layers because the CuO$_2$ planes contribute little to those behaviors (particularly resistivity) above $T_\textrm{c}$ because of disappearing hole-dopant ion separation.

For both cuprates and iron-based superconductors, particle-hole-asymmetric superconductivity domes have been revealed in phase diagrams. We believe that it originates from particle-hole-asymmetric band offsets of different reservoir layers with regard to CuO$_2$ that make up the superconductor, because the Fermi level of the undoped charge reservoir layers and the mid-gap position of the Mott insulator CuO$_2$ are usually different. In the case that the band offset does not allow enough charge transfer, superconductivity dome may not emerge.

Finally, we would like to point out that the monolayer oxide films studied here might be T-CuO, whose ground state is also a Mott insulator and similar to CuO$_2$ \cite{Moser2014angle}. As discussed above, whether it is CuO$_2$ or T-CuO does not affect the modulation-doping scheme: as long as a material can form an ideal heterostructure with it and their respective electronic structures satisfy the criterion for band bending, the system may superconduct. The FeSe/SrTiO$_3$ and LaAlO$_3$/SrTiO$_3$ heterostructures represent two good examples \cite{qing2012interface, reyren2007superconducting}.
\\
\newline
\newline
\textbf{4 Summary}

Our model is based on the modulation-doping scheme that is in the framework of rigid band model and has been proved extremely successful in semiconductor physics since its development in late 1970$^\prime$s \cite{dingle1978electron}. Although this rigid-band model based scheme is simple, we have shown that it could capture the essential features of Bi-2212 for both normal and superconducting states in the entire phase diagram. The model may also work for other cuprates and iron-based superconductors in terms of the same doping-charge-transfer mechanism. To justify the model, here we propose to employ electrical field effect (e.g.\ by ionic liquids) to tune the copper-oxide layers on non-superconducting substrates (such as SrTiO$_3$ (001)) superconducting. If the model is confirmed, it would point to an explicit direction for discovering new high-\emph{T}$_c$ materials, as has been done for the quantum Hall effect with various semiconductor heterostructures. Last but not least, the observed nodeless superconductivity implies that the pairing might be mediated by an attractive interaction in $k$ space, namely the electron-phonon interaction in conventional superconductors, but with enhanced Debye temperature and electron-phonon coupling \cite{qing2012interface}. Therefore, exploring the isotope effect of $T_\textrm{c}$ will help unravel the pairing mechanism. The successful growth of CuO$_2$ monolayer films, as reported here, paves a solid base for future study. \par

\begin{acknowledgments}
The work was financially supported by National Science Foundation, Ministry of Science and Technology and Ministry of Education of China. The work at Brookhaven National Laboratory was supported by the Office of Basic Energy Sciences, U.S. Department of Energy, under Contract No. DE-SC00112704.
\end{acknowledgments}

\noindent \textbf{Conflict of interest} The authors declare that they have no conflict of interest.


\begin{thebibliography}{32}%
\makeatletter
\providecommand \@ifxundefined [1]{%
 \@ifx{#1\undefined}
}%
\providecommand \@ifnum [1]{%
 \ifnum #1\expandafter \@firstoftwo
 \else \expandafter \@secondoftwo
 \fi
}%
\providecommand \@ifx [1]{%
 \ifx #1\expandafter \@firstoftwo
 \else \expandafter \@secondoftwo
 \fi
}%
\providecommand \natexlab [1]{#1}%
\providecommand \enquote  [1]{``#1''}%
\providecommand \bibnamefont  [1]{#1}%
\providecommand \bibfnamefont [1]{#1}%
\providecommand \citenamefont [1]{#1}%
\providecommand \href@noop [0]{\@secondoftwo}%
\providecommand \href [0]{\begingroup \@sanitize@url \@href}%
\providecommand \@href[1]{\@@startlink{#1}\@@href}%
\providecommand \@@href[1]{\endgroup#1\@@endlink}%
\providecommand \@sanitize@url [0]{\catcode `\\12\catcode `\$12\catcode
  `\&12\catcode `\#12\catcode `\^12\catcode `\_12\catcode `\%12\relax}%
\providecommand \@@startlink[1]{}%
\providecommand \@@endlink[0]{}%
\providecommand \url  [0]{\begingroup\@sanitize@url \@url }%
\providecommand \@url [1]{\endgroup\@href {#1}{\urlprefix }}%
\providecommand \urlprefix  [0]{URL }%
\providecommand \Eprint [0]{\href }%
\providecommand \doibase [0]{http://dx.doi.org/}%
\providecommand \selectlanguage [0]{\@gobble}%
\providecommand \bibinfo  [0]{\@secondoftwo}%
\providecommand \bibfield  [0]{\@secondoftwo}%
\providecommand \translation [1]{[#1]}%
\providecommand \BibitemOpen [0]{}%
\providecommand \bibitemStop [0]{}%
\providecommand \bibitemNoStop [0]{.\EOS\space}%
\providecommand \EOS [0]{\spacefactor3000\relax}%
\providecommand \BibitemShut  [1]{\csname bibitem#1\endcsname}%
\let\auto@bib@innerbib\@empty
\bibitem [{\citenamefont {Bednorz}\ and\ \citenamefont
  {M{\"u}ller}(1986)}]{bednorz1986possible}%
  \BibitemOpen
  \bibfield  {author} {\bibinfo {author} {\bibfnamefont {J.~G.}\ \bibnamefont
  {Bednorz}}\ and\ \bibinfo {author} {\bibfnamefont {K.~A.}\ \bibnamefont
  {M{\"u}ller}},\ }\href {\doibase 10.1007/BF01303701} {\bibfield  {journal}
  {\bibinfo  {journal} {Z. Phys. B}\ }\textbf {\bibinfo {volume} {64}},\
  \bibinfo {pages} {189} (\bibinfo {year} {1986})}\BibitemShut {NoStop}%
\bibitem [{\citenamefont {Wang}\ \emph {et~al.}(2012)\citenamefont {Wang},
  \citenamefont {Li}, \citenamefont {Zhang}, \citenamefont {Zhang},
  \citenamefont {Zhang}, \citenamefont {LI}, \citenamefont {Ding},
  \citenamefont {Ou}, \citenamefont {Deng}, \citenamefont {Chang},
  \citenamefont {Wen}, \citenamefont {Song}, \citenamefont {He}, \citenamefont
  {Jia}, \citenamefont {Ji}, \citenamefont {Wang}, \citenamefont {Wang},
  \citenamefont {Chen}, \citenamefont {Ma},\ and\ \citenamefont
  {Xue}}]{qing2012interface}%
  \BibitemOpen
  \bibfield  {author} {\bibinfo {author} {\bibfnamefont {Q.~Y.}\ \bibnamefont
  {Wang}}, \bibinfo {author} {\bibfnamefont {Z.}~\bibnamefont {Li}}, \bibinfo
  {author} {\bibfnamefont {W.~H.}\ \bibnamefont {Zhang}}, \bibinfo {author}
  {\bibfnamefont {Z.~C.}\ \bibnamefont {Zhang}}, \bibinfo {author}
  {\bibfnamefont {J.~S.}\ \bibnamefont {Zhang}}, \bibinfo {author}
  {\bibfnamefont {W.}~\bibnamefont {LI}}, \bibinfo {author} {\bibfnamefont
  {H.}~\bibnamefont {Ding}}, \bibinfo {author} {\bibfnamefont {Y.~B.}\
  \bibnamefont {Ou}}, \bibinfo {author} {\bibfnamefont {P.}~\bibnamefont
  {Deng}}, \bibinfo {author} {\bibfnamefont {K.}~\bibnamefont {Chang}},
  \bibinfo {author} {\bibfnamefont {J.}~\bibnamefont {Wen}}, \bibinfo {author}
  {\bibfnamefont {C.~L.}\ \bibnamefont {Song}}, \bibinfo {author}
  {\bibfnamefont {K.}~\bibnamefont {He}}, \bibinfo {author} {\bibfnamefont
  {J.~F.}\ \bibnamefont {Jia}}, \bibinfo {author} {\bibfnamefont {S.~H.}\
  \bibnamefont {Ji}}, \bibinfo {author} {\bibfnamefont {Y.~Y.}\ \bibnamefont
  {Wang}}, \bibinfo {author} {\bibfnamefont {L.~L.}\ \bibnamefont {Wang}},
  \bibinfo {author} {\bibfnamefont {X.}~\bibnamefont {Chen}}, \bibinfo {author}
  {\bibfnamefont {X.~C.}\ \bibnamefont {Ma}}, \ and\ \bibinfo {author}
  {\bibfnamefont {Q.~K.}\ \bibnamefont {Xue}},\ }\href {\doibase
  10.1088/0256-307X/29/3/037402} {\bibfield  {journal} {\bibinfo  {journal}
  {Chin. Phys. Lett.}\ }\textbf {\bibinfo {volume} {29}},\ \bibinfo {pages}
  {037402} (\bibinfo {year} {2012})}\BibitemShut {NoStop}%
\bibitem [{\citenamefont {Lv}\ \emph {et~al.}(2015)\citenamefont {Lv},
  \citenamefont {Wang}, \citenamefont {Peng}, \citenamefont {Ding},
  \citenamefont {Wang}, \citenamefont {Wang}, \citenamefont {He}, \citenamefont
  {Ji}, \citenamefont {Zhong}, \citenamefont {Schneeloch}, \citenamefont {Gu},
  \citenamefont {Song}, \citenamefont {Ma},\ and\ \citenamefont
  {Xue}}]{Lv2015mapping}%
  \BibitemOpen
  \bibfield  {author} {\bibinfo {author} {\bibfnamefont {Y.~F.}\ \bibnamefont
  {Lv}}, \bibinfo {author} {\bibfnamefont {W.~L.}\ \bibnamefont {Wang}},
  \bibinfo {author} {\bibfnamefont {J.~P.}\ \bibnamefont {Peng}}, \bibinfo
  {author} {\bibfnamefont {H.}~\bibnamefont {Ding}}, \bibinfo {author}
  {\bibfnamefont {Y.}~\bibnamefont {Wang}}, \bibinfo {author} {\bibfnamefont
  {L.}~\bibnamefont {Wang}}, \bibinfo {author} {\bibfnamefont {K.}~\bibnamefont
  {He}}, \bibinfo {author} {\bibfnamefont {S.~H.}\ \bibnamefont {Ji}}, \bibinfo
  {author} {\bibfnamefont {R.}~\bibnamefont {Zhong}}, \bibinfo {author}
  {\bibfnamefont {J.}~\bibnamefont {Schneeloch}}, \bibinfo {author}
  {\bibfnamefont {G.~D.}\ \bibnamefont {Gu}}, \bibinfo {author} {\bibfnamefont
  {C.~L.}\ \bibnamefont {Song}}, \bibinfo {author} {\bibfnamefont {X.~C.}\
  \bibnamefont {Ma}}, \ and\ \bibinfo {author} {\bibfnamefont {Q.~K.}\
  \bibnamefont {Xue}},\ }\href {\doibase 10.1103/PhysRevLett.115.237002}
  {\bibfield  {journal} {\bibinfo  {journal} {Phys. Rev. Lett.}\ }\textbf
  {\bibinfo {volume} {115}},\ \bibinfo {pages} {237002} (\bibinfo {year}
  {2015})}\BibitemShut {NoStop}%
\bibitem [{\citenamefont {Lv}\ \emph {et~al.}(2016)\citenamefont {Lv},
  \citenamefont {Wang}, \citenamefont {Ding}, \citenamefont {Wang},
  \citenamefont {Ding}, \citenamefont {Zhong}, \citenamefont {Schneeloch},
  \citenamefont {Gu}, \citenamefont {Wang}, \citenamefont {He}, \citenamefont
  {Ji}, \citenamefont {Zhao}, \citenamefont {Zhou}, \citenamefont {Song},
  \citenamefont {Ma},\ and\ \citenamefont {Xue}}]{Lv2016electronic}%
  \BibitemOpen
  \bibfield  {author} {\bibinfo {author} {\bibfnamefont {Y.~F.}\ \bibnamefont
  {Lv}}, \bibinfo {author} {\bibfnamefont {W.~L.}\ \bibnamefont {Wang}},
  \bibinfo {author} {\bibfnamefont {H.}~\bibnamefont {Ding}}, \bibinfo {author}
  {\bibfnamefont {Y.}~\bibnamefont {Wang}}, \bibinfo {author} {\bibfnamefont
  {Y.}~\bibnamefont {Ding}}, \bibinfo {author} {\bibfnamefont {R.}~\bibnamefont
  {Zhong}}, \bibinfo {author} {\bibfnamefont {J.}~\bibnamefont {Schneeloch}},
  \bibinfo {author} {\bibfnamefont {G.~D.}\ \bibnamefont {Gu}}, \bibinfo
  {author} {\bibfnamefont {L.}~\bibnamefont {Wang}}, \bibinfo {author}
  {\bibfnamefont {K.}~\bibnamefont {He}}, \bibinfo {author} {\bibfnamefont
  {S.-H.}\ \bibnamefont {Ji}}, \bibinfo {author} {\bibfnamefont
  {L.}~\bibnamefont {Zhao}}, \bibinfo {author} {\bibfnamefont {X.~J.}\
  \bibnamefont {Zhou}}, \bibinfo {author} {\bibfnamefont {C.~L.}\ \bibnamefont
  {Song}}, \bibinfo {author} {\bibfnamefont {X.~C.}\ \bibnamefont {Ma}}, \ and\
  \bibinfo {author} {\bibfnamefont {Q.~K.}\ \bibnamefont {Xue}},\ }\href
  {\doibase 10.1103/PhysRevB.93.140504} {\bibfield  {journal} {\bibinfo
  {journal} {Phys. Rev. B}\ }\textbf {\bibinfo {volume} {93}},\ \bibinfo
  {pages} {140504} (\bibinfo {year} {2016})}\BibitemShut {NoStop}%
\bibitem [{\citenamefont {Siemons}\ \emph {et~al.}(2009)\citenamefont
  {Siemons}, \citenamefont {Koster}, \citenamefont {Blank}, \citenamefont
  {Hammond}, \citenamefont {Geballe},\ and\ \citenamefont
  {Beasley}}]{Siemons2009tetragonal}%
  \BibitemOpen
  \bibfield  {author} {\bibinfo {author} {\bibfnamefont {W.}~\bibnamefont
  {Siemons}}, \bibinfo {author} {\bibfnamefont {G.}~\bibnamefont {Koster}},
  \bibinfo {author} {\bibfnamefont {D.~H.~A.}\ \bibnamefont {Blank}}, \bibinfo
  {author} {\bibfnamefont {R.~H.}\ \bibnamefont {Hammond}}, \bibinfo {author}
  {\bibfnamefont {T.~H.}\ \bibnamefont {Geballe}}, \ and\ \bibinfo {author}
  {\bibfnamefont {M.~R.}\ \bibnamefont {Beasley}},\ }\href {\doibase
  10.1103/PhysRevB.79.195122} {\bibfield  {journal} {\bibinfo  {journal} {Phys.
  Rev. B}\ }\textbf {\bibinfo {volume} {79}},\ \bibinfo {pages} {195122}
  (\bibinfo {year} {2009})}\BibitemShut {NoStop}%
\bibitem [{\citenamefont {Moser}\ \emph {et~al.}(2014)\citenamefont {Moser},
  \citenamefont {Moreschini}, \citenamefont {Yang}, \citenamefont {Innocenti},
  \citenamefont {Fuchs}, \citenamefont {Hansen}, \citenamefont {Chang},
  \citenamefont {Kim}, \citenamefont {Walter}, \citenamefont {Bostwick},
  \citenamefont {Rotenberg}, \citenamefont {Mila},\ and\ \citenamefont
  {Grioni}}]{Moser2014angle}%
  \BibitemOpen
  \bibfield  {author} {\bibinfo {author} {\bibfnamefont {S.}~\bibnamefont
  {Moser}}, \bibinfo {author} {\bibfnamefont {L.}~\bibnamefont {Moreschini}},
  \bibinfo {author} {\bibfnamefont {H.-Y.}\ \bibnamefont {Yang}}, \bibinfo
  {author} {\bibfnamefont {D.}~\bibnamefont {Innocenti}}, \bibinfo {author}
  {\bibfnamefont {F.}~\bibnamefont {Fuchs}}, \bibinfo {author} {\bibfnamefont
  {N.~H.}\ \bibnamefont {Hansen}}, \bibinfo {author} {\bibfnamefont {Y.~J.}\
  \bibnamefont {Chang}}, \bibinfo {author} {\bibfnamefont {K.~S.}\ \bibnamefont
  {Kim}}, \bibinfo {author} {\bibfnamefont {A.~L.}\ \bibnamefont {Walter}},
  \bibinfo {author} {\bibfnamefont {A.}~\bibnamefont {Bostwick}}, \bibinfo
  {author} {\bibfnamefont {E.}~\bibnamefont {Rotenberg}}, \bibinfo {author}
  {\bibfnamefont {F.}~\bibnamefont {Mila}}, \ and\ \bibinfo {author}
  {\bibfnamefont {M.}~\bibnamefont {Grioni}},\ }\href {\doibase
  10.1103/PhysRevLett.113.187001} {\bibfield  {journal} {\bibinfo  {journal}
  {Phys. Rev. Lett.}\ }\textbf {\bibinfo {volume} {113}},\ \bibinfo {pages}
  {187001} (\bibinfo {year} {2014})}\BibitemShut {NoStop}%
\bibitem [{\citenamefont {Adolphs}\ \emph {et~al.}(2016)\citenamefont
  {Adolphs}, \citenamefont {Moser}, \citenamefont {Sawatzky},\ and\
  \citenamefont {Berciu}}]{Adolphs2016non}%
  \BibitemOpen
  \bibfield  {author} {\bibinfo {author} {\bibfnamefont {C.~P.~J.}\
  \bibnamefont {Adolphs}}, \bibinfo {author} {\bibfnamefont {S.}~\bibnamefont
  {Moser}}, \bibinfo {author} {\bibfnamefont {G.~A.}\ \bibnamefont {Sawatzky}},
  \ and\ \bibinfo {author} {\bibfnamefont {M.}~\bibnamefont {Berciu}},\ }\href
  {\doibase 10.1103/PhysRevLett.116.087002} {\bibfield  {journal} {\bibinfo
  {journal} {Phys. Rev. Lett.}\ }\textbf {\bibinfo {volume} {116}},\ \bibinfo
  {pages} {087002} (\bibinfo {year} {2016})}\BibitemShut {NoStop}%
\bibitem [{\citenamefont {Fischer}\ \emph {et~al.}(2007)\citenamefont
  {Fischer}, \citenamefont {Kugler}, \citenamefont {Maggio-Aprile},
  \citenamefont {Berthod},\ and\ \citenamefont {Renner}}]{fischer2007scanning}%
  \BibitemOpen
  \bibfield  {author} {\bibinfo {author} {\bibfnamefont {{\O}.}~\bibnamefont
  {Fischer}}, \bibinfo {author} {\bibfnamefont {M.}~\bibnamefont {Kugler}},
  \bibinfo {author} {\bibfnamefont {I.}~\bibnamefont {Maggio-Aprile}}, \bibinfo
  {author} {\bibfnamefont {C.}~\bibnamefont {Berthod}}, \ and\ \bibinfo
  {author} {\bibfnamefont {C.}~\bibnamefont {Renner}},\ }\href {\doibase
  10.1103/RevModPhys.79.353} {\bibfield  {journal} {\bibinfo  {journal} {Rev.
  Mod. Phys.}\ }\textbf {\bibinfo {volume} {79}},\ \bibinfo {pages} {353}
  (\bibinfo {year} {2007})}\BibitemShut {NoStop}%
\bibitem [{\citenamefont {Ye}\ \emph {et~al.}(2013)\citenamefont {Ye},
  \citenamefont {Cai}, \citenamefont {Yu}, \citenamefont {Zhou}, \citenamefont
  {Ruan}, \citenamefont {Liu}, \citenamefont {Jin},\ and\ \citenamefont
  {Wang}}]{ye2013visualizing}%
  \BibitemOpen
  \bibfield  {author} {\bibinfo {author} {\bibfnamefont {C.}~\bibnamefont
  {Ye}}, \bibinfo {author} {\bibfnamefont {P.}~\bibnamefont {Cai}}, \bibinfo
  {author} {\bibfnamefont {R.}~\bibnamefont {Yu}}, \bibinfo {author}
  {\bibfnamefont {X.}~\bibnamefont {Zhou}}, \bibinfo {author} {\bibfnamefont
  {W.}~\bibnamefont {Ruan}}, \bibinfo {author} {\bibfnamefont {Q.}~\bibnamefont
  {Liu}}, \bibinfo {author} {\bibfnamefont {C.}~\bibnamefont {Jin}}, \ and\
  \bibinfo {author} {\bibfnamefont {Y.}~\bibnamefont {Wang}},\ }\href {\doibase
  10.1038/ncomms2369} {\bibfield  {journal} {\bibinfo  {journal} {Nat.
  Commun.}\ }\textbf {\bibinfo {volume} {4}},\ \bibinfo {pages} {1365}
  (\bibinfo {year} {2013})}\BibitemShut {NoStop}%
\bibitem [{\citenamefont {Zaanen}\ \emph {et~al.}(1985)\citenamefont {Zaanen},
  \citenamefont {Sawatzky},\ and\ \citenamefont {Allen}}]{PhysRevLett.55.418}%
  \BibitemOpen
  \bibfield  {author} {\bibinfo {author} {\bibfnamefont {J.}~\bibnamefont
  {Zaanen}}, \bibinfo {author} {\bibfnamefont {G.~A.}\ \bibnamefont
  {Sawatzky}}, \ and\ \bibinfo {author} {\bibfnamefont {J.~W.}\ \bibnamefont
  {Allen}},\ }\href {\doibase 10.1103/PhysRevLett.55.418} {\bibfield  {journal}
  {\bibinfo  {journal} {Phys. Rev. Lett.}\ }\textbf {\bibinfo {volume} {55}},\
  \bibinfo {pages} {418} (\bibinfo {year} {1985})}\BibitemShut {NoStop}%
\bibitem [{\citenamefont {Zhang}\ and\ \citenamefont
  {Rice}(1988)}]{PhysRevB.37.3759}%
  \BibitemOpen
  \bibfield  {author} {\bibinfo {author} {\bibfnamefont {F.~C.}\ \bibnamefont
  {Zhang}}\ and\ \bibinfo {author} {\bibfnamefont {T.~M.}\ \bibnamefont
  {Rice}},\ }\href {\doibase 10.1103/PhysRevB.37.3759} {\bibfield  {journal}
  {\bibinfo  {journal} {Phys. Rev. B}\ }\textbf {\bibinfo {volume} {37}},\
  \bibinfo {pages} {3759} (\bibinfo {year} {1988})}\BibitemShut {NoStop}%
\bibitem [{\citenamefont {Kastner}\ \emph {et~al.}(1998)\citenamefont
  {Kastner}, \citenamefont {Birgeneau}, \citenamefont {Shirane},\ and\
  \citenamefont {Endoh}}]{RevModPhys.70.897}%
  \BibitemOpen
  \bibfield  {author} {\bibinfo {author} {\bibfnamefont {M.~A.}\ \bibnamefont
  {Kastner}}, \bibinfo {author} {\bibfnamefont {R.~J.}\ \bibnamefont
  {Birgeneau}}, \bibinfo {author} {\bibfnamefont {G.}~\bibnamefont {Shirane}},
  \ and\ \bibinfo {author} {\bibfnamefont {Y.}~\bibnamefont {Endoh}},\ }\href
  {\doibase 10.1103/RevModPhys.70.897} {\bibfield  {journal} {\bibinfo
  {journal} {Rev. Mod. Phys.}\ }\textbf {\bibinfo {volume} {70}},\ \bibinfo
  {pages} {897} (\bibinfo {year} {1998})}\BibitemShut {NoStop}%
\bibitem [{\citenamefont {Hashimoto}\ \emph {et~al.}(2014)\citenamefont
  {Hashimoto}, \citenamefont {Vishik}, \citenamefont {He}, \citenamefont
  {Devereaux},\ and\ \citenamefont {Shen}}]{hashimoto2014energy}%
  \BibitemOpen
  \bibfield  {author} {\bibinfo {author} {\bibfnamefont {M.}~\bibnamefont
  {Hashimoto}}, \bibinfo {author} {\bibfnamefont {I.~M.}\ \bibnamefont
  {Vishik}}, \bibinfo {author} {\bibfnamefont {R.-H.}\ \bibnamefont {He}},
  \bibinfo {author} {\bibfnamefont {T.~P.}\ \bibnamefont {Devereaux}}, \ and\
  \bibinfo {author} {\bibfnamefont {Z.-X.}\ \bibnamefont {Shen}},\ }\href
  {\doibase 10.1038/nphys3009} {\bibfield  {journal} {\bibinfo  {journal} {Nat.
  Phys.}\ }\textbf {\bibinfo {volume} {10}},\ \bibinfo {pages} {483} (\bibinfo
  {year} {2014})}\BibitemShut {NoStop}%
\bibitem [{\citenamefont {Pushp}\ \emph {et~al.}(2009)\citenamefont {Pushp},
  \citenamefont {Parker}, \citenamefont {Pasupathy}, \citenamefont {Gomes},
  \citenamefont {Ono}, \citenamefont {Wen}, \citenamefont {Xu}, \citenamefont
  {Gu},\ and\ \citenamefont {Yazdani}}]{pushp2009extending}%
  \BibitemOpen
  \bibfield  {author} {\bibinfo {author} {\bibfnamefont {A.}~\bibnamefont
  {Pushp}}, \bibinfo {author} {\bibfnamefont {C.~V.}\ \bibnamefont {Parker}},
  \bibinfo {author} {\bibfnamefont {A.~N.}\ \bibnamefont {Pasupathy}}, \bibinfo
  {author} {\bibfnamefont {K.~K.}\ \bibnamefont {Gomes}}, \bibinfo {author}
  {\bibfnamefont {S.}~\bibnamefont {Ono}}, \bibinfo {author} {\bibfnamefont
  {J.}~\bibnamefont {Wen}}, \bibinfo {author} {\bibfnamefont {Z.}~\bibnamefont
  {Xu}}, \bibinfo {author} {\bibfnamefont {G.}~\bibnamefont {Gu}}, \ and\
  \bibinfo {author} {\bibfnamefont {A.}~\bibnamefont {Yazdani}},\ }\href
  {\doibase 10.1126/science.1174338} {\bibfield  {journal} {\bibinfo  {journal}
  {Science}\ }\textbf {\bibinfo {volume} {324}},\ \bibinfo {pages} {1689}
  (\bibinfo {year} {2009})}\BibitemShut {NoStop}%
\bibitem [{\citenamefont {Alldredge}\ \emph {et~al.}(2008)\citenamefont
  {Alldredge}, \citenamefont {Lee}, \citenamefont {McElroy}, \citenamefont
  {Wang}, \citenamefont {Fujita}, \citenamefont {Kohsaka}, \citenamefont
  {Taylor}, \citenamefont {Eisaki}, \citenamefont {Uchida}, \citenamefont
  {Hirschfeld},\ and\ \citenamefont {Davis}}]{alldredge2008evolution}%
  \BibitemOpen
  \bibfield  {author} {\bibinfo {author} {\bibfnamefont {J.~W.}\ \bibnamefont
  {Alldredge}}, \bibinfo {author} {\bibfnamefont {J.}~\bibnamefont {Lee}},
  \bibinfo {author} {\bibfnamefont {K.}~\bibnamefont {McElroy}}, \bibinfo
  {author} {\bibfnamefont {M.}~\bibnamefont {Wang}}, \bibinfo {author}
  {\bibfnamefont {K.}~\bibnamefont {Fujita}}, \bibinfo {author} {\bibfnamefont
  {Y.}~\bibnamefont {Kohsaka}}, \bibinfo {author} {\bibfnamefont
  {C.}~\bibnamefont {Taylor}}, \bibinfo {author} {\bibfnamefont
  {H.}~\bibnamefont {Eisaki}}, \bibinfo {author} {\bibfnamefont
  {S.}~\bibnamefont {Uchida}}, \bibinfo {author} {\bibfnamefont {P.~J.}\
  \bibnamefont {Hirschfeld}}, \ and\ \bibinfo {author} {\bibfnamefont {J.~C.}\
  \bibnamefont {Davis}},\ }\href {\doibase 10.1038/nphys917} {\bibfield
  {journal} {\bibinfo  {journal} {Nat. Phys.}\ }\textbf {\bibinfo {volume}
  {4}},\ \bibinfo {pages} {319} (\bibinfo {year} {2008})}\BibitemShut {NoStop}%
\bibitem [{\citenamefont {Pan}\ \emph {et~al.}(2001)\citenamefont {Pan},
  \citenamefont {O'neal}, \citenamefont {Badzey}, \citenamefont {Chamon},
  \citenamefont {Ding}, \citenamefont {Engelbrecht}, \citenamefont {Wang},
  \citenamefont {Eisaki}, \citenamefont {Uchida}, \citenamefont {Gupta},\ and\
  \citenamefont {Davis}}]{pan2001microscopic}%
  \BibitemOpen
  \bibfield  {author} {\bibinfo {author} {\bibfnamefont {S.~H.}\ \bibnamefont
  {Pan}}, \bibinfo {author} {\bibfnamefont {J.~P.}\ \bibnamefont {O'neal}},
  \bibinfo {author} {\bibfnamefont {R.~L.}\ \bibnamefont {Badzey}}, \bibinfo
  {author} {\bibfnamefont {C.}~\bibnamefont {Chamon}}, \bibinfo {author}
  {\bibfnamefont {H.}~\bibnamefont {Ding}}, \bibinfo {author} {\bibfnamefont
  {J.~R.}\ \bibnamefont {Engelbrecht}}, \bibinfo {author} {\bibfnamefont
  {Z.}~\bibnamefont {Wang}}, \bibinfo {author} {\bibfnamefont {H.}~\bibnamefont
  {Eisaki}}, \bibinfo {author} {\bibfnamefont {S.}~\bibnamefont {Uchida}},
  \bibinfo {author} {\bibfnamefont {A.~K.}\ \bibnamefont {Gupta}}, \ and\
  \bibinfo {author} {\bibfnamefont {J.~C.}\ \bibnamefont {Davis}},\ }\href@noop
  {} {\bibfield  {journal} {\bibinfo  {journal} {Nature}\ }\textbf {\bibinfo
  {volume} {413}},\ \bibinfo {pages} {282} (\bibinfo {year}
  {2001})}\BibitemShut {NoStop}%
\bibitem [{\citenamefont {McElroy}\ \emph {et~al.}(2005)\citenamefont
  {McElroy}, \citenamefont {Lee}, \citenamefont {Slezak}, \citenamefont {Lee},
  \citenamefont {Eisaki}, \citenamefont {Uchida},\ and\ \citenamefont
  {Davis}}]{mcelroy2005atomic}%
  \BibitemOpen
  \bibfield  {author} {\bibinfo {author} {\bibfnamefont {K.}~\bibnamefont
  {McElroy}}, \bibinfo {author} {\bibfnamefont {J.}~\bibnamefont {Lee}},
  \bibinfo {author} {\bibfnamefont {J.}~\bibnamefont {Slezak}}, \bibinfo
  {author} {\bibfnamefont {D.-H.}\ \bibnamefont {Lee}}, \bibinfo {author}
  {\bibfnamefont {H.}~\bibnamefont {Eisaki}}, \bibinfo {author} {\bibfnamefont
  {S.}~\bibnamefont {Uchida}}, \ and\ \bibinfo {author} {\bibfnamefont
  {J.}~\bibnamefont {Davis}},\ }\href {\doibase 10.1126/science.1113095}
  {\bibfield  {journal} {\bibinfo  {journal} {Science}\ }\textbf {\bibinfo
  {volume} {309}},\ \bibinfo {pages} {1048} (\bibinfo {year}
  {2005})}\BibitemShut {NoStop}%
\bibitem [{\citenamefont {H{\"u}fner}\ \emph {et~al.}(2008)\citenamefont
  {H{\"u}fner}, \citenamefont {Hossain}, \citenamefont {Damascelli},\ and\
  \citenamefont {Sawatzky}}]{hufner2008two}%
  \BibitemOpen
  \bibfield  {author} {\bibinfo {author} {\bibfnamefont {S.}~\bibnamefont
  {H{\"u}fner}}, \bibinfo {author} {\bibfnamefont {M.}~\bibnamefont {Hossain}},
  \bibinfo {author} {\bibfnamefont {A.}~\bibnamefont {Damascelli}}, \ and\
  \bibinfo {author} {\bibfnamefont {G.}~\bibnamefont {Sawatzky}},\ }\href
  {\doibase 10.1088/0034-4885/71/6/062501} {\bibfield  {journal} {\bibinfo
  {journal} {Rep. Prog. Phys.}\ }\textbf {\bibinfo {volume} {71}},\ \bibinfo
  {pages} {062501} (\bibinfo {year} {2008})}\BibitemShut {NoStop}%
\bibitem [{\citenamefont {Kondo}\ \emph {et~al.}(2009)\citenamefont {Kondo},
  \citenamefont {Khasanov}, \citenamefont {Takeuchi}, \citenamefont
  {Schmalian},\ and\ \citenamefont {Kaminski}}]{kondo2009competition}%
  \BibitemOpen
  \bibfield  {author} {\bibinfo {author} {\bibfnamefont {T.}~\bibnamefont
  {Kondo}}, \bibinfo {author} {\bibfnamefont {R.}~\bibnamefont {Khasanov}},
  \bibinfo {author} {\bibfnamefont {T.}~\bibnamefont {Takeuchi}}, \bibinfo
  {author} {\bibfnamefont {J.}~\bibnamefont {Schmalian}}, \ and\ \bibinfo
  {author} {\bibfnamefont {A.}~\bibnamefont {Kaminski}},\ }\href {\doibase
  10.1038/nature07644} {\bibfield  {journal} {\bibinfo  {journal} {Nature}\
  }\textbf {\bibinfo {volume} {457}},\ \bibinfo {pages} {296} (\bibinfo {year}
  {2009})}\BibitemShut {NoStop}%
\bibitem [{\citenamefont {Chakravarty}\ \emph {et~al.}(2001)\citenamefont
  {Chakravarty}, \citenamefont {Laughlin}, \citenamefont {Morr},\ and\
  \citenamefont {Nayak}}]{chakravarty2001hidden}%
  \BibitemOpen
  \bibfield  {author} {\bibinfo {author} {\bibfnamefont {S.}~\bibnamefont
  {Chakravarty}}, \bibinfo {author} {\bibfnamefont {R.}~\bibnamefont
  {Laughlin}}, \bibinfo {author} {\bibfnamefont {D.~K.}\ \bibnamefont {Morr}},
  \ and\ \bibinfo {author} {\bibfnamefont {C.}~\bibnamefont {Nayak}},\ }\href
  {\doibase 10.1103/PhysRevB.63.094503} {\bibfield  {journal} {\bibinfo
  {journal} {Phys. Rev. B}\ }\textbf {\bibinfo {volume} {63}},\ \bibinfo
  {pages} {094503} (\bibinfo {year} {2001})}\BibitemShut {NoStop}%
\bibitem [{\citenamefont {Yazdani}\ \emph {et~al.}(1997)\citenamefont
  {Yazdani}, \citenamefont {Jones}, \citenamefont {Lutz}, \citenamefont
  {Crommie},\ and\ \citenamefont {Eigler}}]{yazdani1997probing}%
  \BibitemOpen
  \bibfield  {author} {\bibinfo {author} {\bibfnamefont {A.}~\bibnamefont
  {Yazdani}}, \bibinfo {author} {\bibfnamefont {B.~A.}\ \bibnamefont {Jones}},
  \bibinfo {author} {\bibfnamefont {C.~P.}\ \bibnamefont {Lutz}}, \bibinfo
  {author} {\bibfnamefont {M.~F.}\ \bibnamefont {Crommie}}, \ and\ \bibinfo
  {author} {\bibfnamefont {D.~M.}\ \bibnamefont {Eigler}},\ }\href {\doibase
  10.1126/science.275.5307.1767} {\bibfield  {journal} {\bibinfo  {journal}
  {Science}\ }\textbf {\bibinfo {volume} {275}},\ \bibinfo {pages} {1767}
  (\bibinfo {year} {1997})}\BibitemShut {NoStop}%
\bibitem [{\citenamefont {Balatsky}\ \emph {et~al.}(2006)\citenamefont
  {Balatsky}, \citenamefont {Vekhter},\ and\ \citenamefont
  {Zhu}}]{balatsky2006impurity}%
  \BibitemOpen
  \bibfield  {author} {\bibinfo {author} {\bibfnamefont {A.}~\bibnamefont
  {Balatsky}}, \bibinfo {author} {\bibfnamefont {I.}~\bibnamefont {Vekhter}}, \
  and\ \bibinfo {author} {\bibfnamefont {J.-X.}\ \bibnamefont {Zhu}},\ }\href
  {\doibase http://dx.doi.org/10.1103/RevModPhys.78.373} {\bibfield  {journal}
  {\bibinfo  {journal} {Rev. Mod. Phys.}\ }\textbf {\bibinfo {volume} {78}},\
  \bibinfo {pages} {373} (\bibinfo {year} {2006})}\BibitemShut {NoStop}%
\bibitem [{\citenamefont {Kim}\ \emph {et~al.}(2014)\citenamefont {Kim},
  \citenamefont {Krupin}, \citenamefont {Denlinger}, \citenamefont {Bostwick},
  \citenamefont {Rotenberg}, \citenamefont {Zhao}, \citenamefont {Mitchell},
  \citenamefont {Allen},\ and\ \citenamefont {Kim}}]{kim2014fermi}%
  \BibitemOpen
  \bibfield  {author} {\bibinfo {author} {\bibfnamefont {Y.}~\bibnamefont
  {Kim}}, \bibinfo {author} {\bibfnamefont {O.}~\bibnamefont {Krupin}},
  \bibinfo {author} {\bibfnamefont {J.}~\bibnamefont {Denlinger}}, \bibinfo
  {author} {\bibfnamefont {A.}~\bibnamefont {Bostwick}}, \bibinfo {author}
  {\bibfnamefont {E.}~\bibnamefont {Rotenberg}}, \bibinfo {author}
  {\bibfnamefont {Q.}~\bibnamefont {Zhao}}, \bibinfo {author} {\bibfnamefont
  {J.}~\bibnamefont {Mitchell}}, \bibinfo {author} {\bibfnamefont
  {J.}~\bibnamefont {Allen}}, \ and\ \bibinfo {author} {\bibfnamefont
  {B.}~\bibnamefont {Kim}},\ }\href {\doibase 10.1126/science.1251151}
  {\bibfield  {journal} {\bibinfo  {journal} {Science}\ }\textbf {\bibinfo
  {volume} {345}},\ \bibinfo {pages} {187} (\bibinfo {year}
  {2014})}\BibitemShut {NoStop}%
\bibitem [{\citenamefont {Kim}\ \emph {et~al.}(2016)\citenamefont {Kim},
  \citenamefont {Sung}, \citenamefont {Denlinger},\ and\ \citenamefont
  {Kim}}]{kim2016observation}%
  \BibitemOpen
  \bibfield  {author} {\bibinfo {author} {\bibfnamefont {Y.}~\bibnamefont
  {Kim}}, \bibinfo {author} {\bibfnamefont {N.}~\bibnamefont {Sung}}, \bibinfo
  {author} {\bibfnamefont {J.}~\bibnamefont {Denlinger}}, \ and\ \bibinfo
  {author} {\bibfnamefont {B.}~\bibnamefont {Kim}},\ }\href {\doibase
  10.1038/nphys3503} {\bibfield  {journal} {\bibinfo  {journal} {Nat. Phys.}\
  }\textbf {\bibinfo {volume} {12}},\ \bibinfo {pages} {37} (\bibinfo {year}
  {2016})}\BibitemShut {NoStop}%
\bibitem [{\citenamefont {Su}\ \emph {et~al.}(2006)\citenamefont {Su},
  \citenamefont {Luo},\ and\ \citenamefont {Xiang}}]{PhysRevB.73.134510}%
  \BibitemOpen
  \bibfield  {author} {\bibinfo {author} {\bibfnamefont {Y.~H.}\ \bibnamefont
  {Su}}, \bibinfo {author} {\bibfnamefont {H.~G.}\ \bibnamefont {Luo}}, \ and\
  \bibinfo {author} {\bibfnamefont {T.}~\bibnamefont {Xiang}},\ }\href
  {\doibase 10.1103/PhysRevB.73.134510} {\bibfield  {journal} {\bibinfo
  {journal} {Phys. Rev. B}\ }\textbf {\bibinfo {volume} {73}},\ \bibinfo
  {pages} {134510} (\bibinfo {year} {2006})}\BibitemShut {NoStop}%
\bibitem [{\citenamefont {Mishra}\ \emph {et~al.}(2014)\citenamefont {Mishra},
  \citenamefont {Chatterjee}, \citenamefont {Campuzano},\ and\ \citenamefont
  {Norman}}]{mishra2014effect}%
  \BibitemOpen
  \bibfield  {author} {\bibinfo {author} {\bibfnamefont {V.}~\bibnamefont
  {Mishra}}, \bibinfo {author} {\bibfnamefont {U.}~\bibnamefont {Chatterjee}},
  \bibinfo {author} {\bibfnamefont {J.~C.}\ \bibnamefont {Campuzano}}, \ and\
  \bibinfo {author} {\bibfnamefont {M.~R.}\ \bibnamefont {Norman}},\ }\href
  {\doibase 10.1038/nphys2926} {\bibfield  {journal} {\bibinfo  {journal} {Nat.
  Phys.}\ }\textbf {\bibinfo {volume} {10}},\ \bibinfo {pages} {357} (\bibinfo
  {year} {2014})}\BibitemShut {NoStop}%
\bibitem [{\citenamefont {Dingle}\ \emph {et~al.}(1978)\citenamefont {Dingle},
  \citenamefont {St{\"o}rmer}, \citenamefont {Gossard},\ and\ \citenamefont
  {Wiegmann}}]{dingle1978electron}%
  \BibitemOpen
  \bibfield  {author} {\bibinfo {author} {\bibfnamefont {R.}~\bibnamefont
  {Dingle}}, \bibinfo {author} {\bibfnamefont {H.~L.}\ \bibnamefont
  {St{\"o}rmer}}, \bibinfo {author} {\bibfnamefont {A.~C.}\ \bibnamefont
  {Gossard}}, \ and\ \bibinfo {author} {\bibfnamefont {W.}~\bibnamefont
  {Wiegmann}},\ }\href {\doibase dx.doi.org/10.1063/1.90457} {\bibfield
  {journal} {\bibinfo  {journal} {Appl. Phys. Lett.}\ }\textbf {\bibinfo
  {volume} {33}},\ \bibinfo {pages} {665} (\bibinfo {year} {1978})}\BibitemShut
  {NoStop}%
\bibitem [{\citenamefont {Dean}\ \emph {et~al.}(2013)\citenamefont {Dean},
  \citenamefont {Dellea}, \citenamefont {Springell}, \citenamefont
  {Yakhou-Harris}, \citenamefont {Kummer}, \citenamefont {Brookes},
  \citenamefont {Liu}, \citenamefont {Sun}, \citenamefont {Strle},
  \citenamefont {Schmitt}, \citenamefont {Braicovich}, \citenamefont
  {Ghiringhelli}, \citenamefont {Bo\u{z}vi\'{c}},\ and\ \citenamefont
  {Hill}}]{dean2013persistence}%
  \BibitemOpen
  \bibfield  {author} {\bibinfo {author} {\bibfnamefont {M.~P.~M.}\
  \bibnamefont {Dean}}, \bibinfo {author} {\bibfnamefont {G.}~\bibnamefont
  {Dellea}}, \bibinfo {author} {\bibfnamefont {R.}~\bibnamefont {Springell}},
  \bibinfo {author} {\bibfnamefont {F.}~\bibnamefont {Yakhou-Harris}}, \bibinfo
  {author} {\bibfnamefont {K.}~\bibnamefont {Kummer}}, \bibinfo {author}
  {\bibfnamefont {N.}~\bibnamefont {Brookes}}, \bibinfo {author} {\bibfnamefont
  {X.}~\bibnamefont {Liu}}, \bibinfo {author} {\bibfnamefont {Y.}~\bibnamefont
  {Sun}}, \bibinfo {author} {\bibfnamefont {J.}~\bibnamefont {Strle}}, \bibinfo
  {author} {\bibfnamefont {T.}~\bibnamefont {Schmitt}}, \bibinfo {author}
  {\bibfnamefont {L.}~\bibnamefont {Braicovich}}, \bibinfo {author}
  {\bibfnamefont {G.}~\bibnamefont {Ghiringhelli}}, \bibinfo {author}
  {\bibfnamefont {I.}~\bibnamefont {Bo\u{z}vi\'{c}}}, \ and\ \bibinfo {author}
  {\bibfnamefont {P.}~\bibnamefont {Hill}, \bibfnamefont {J}},\ }\href
  {\doibase 10.1038/nmat3723} {\bibfield  {journal} {\bibinfo  {journal} {Nat.
  Mater.}\ }\textbf {\bibinfo {volume} {12}},\ \bibinfo {pages} {1019}
  (\bibinfo {year} {2013})}\BibitemShut {NoStop}%
\bibitem [{\citenamefont {Fan}\ \emph {et~al.}(2015)\citenamefont {Fan},
  \citenamefont {Zhang}, \citenamefont {Liu}, \citenamefont {Yan},
  \citenamefont {Ren}, \citenamefont {Peng}, \citenamefont {Xu}, \citenamefont
  {Xie}, \citenamefont {Hu}, \citenamefont {Zhang},\ and\ \citenamefont
  {L}}]{fan2015plain}%
  \BibitemOpen
  \bibfield  {author} {\bibinfo {author} {\bibfnamefont {Q.}~\bibnamefont
  {Fan}}, \bibinfo {author} {\bibfnamefont {W.~H.}\ \bibnamefont {Zhang}},
  \bibinfo {author} {\bibfnamefont {X.}~\bibnamefont {Liu}}, \bibinfo {author}
  {\bibfnamefont {Y.~J.}\ \bibnamefont {Yan}}, \bibinfo {author} {\bibfnamefont
  {M.~Q.}\ \bibnamefont {Ren}}, \bibinfo {author} {\bibfnamefont
  {R.}~\bibnamefont {Peng}}, \bibinfo {author} {\bibfnamefont {H.~C.}\
  \bibnamefont {Xu}}, \bibinfo {author} {\bibfnamefont {B.~P.}\ \bibnamefont
  {Xie}}, \bibinfo {author} {\bibfnamefont {J.~P.}\ \bibnamefont {Hu}},
  \bibinfo {author} {\bibfnamefont {T.}~\bibnamefont {Zhang}}, \ and\ \bibinfo
  {author} {\bibfnamefont {F.~D.}\ \bibnamefont {L}},\ }\href {\doibase
  10.1038/nphys3450} {\bibfield  {journal} {\bibinfo  {journal} {Nat. Phys.}\
  }\textbf {\bibinfo {volume} {11}},\ \bibinfo {pages} {946} (\bibinfo {year}
  {2015})}\BibitemShut {NoStop}%
\bibitem [{\citenamefont {Keimer}\ \emph {et~al.}(2015)\citenamefont {Keimer},
  \citenamefont {Kivelson}, \citenamefont {Norman}, \citenamefont {Uchida},\
  and\ \citenamefont {Zaanen}}]{keimer2015quantum}%
  \BibitemOpen
  \bibfield  {author} {\bibinfo {author} {\bibfnamefont {B.}~\bibnamefont
  {Keimer}}, \bibinfo {author} {\bibfnamefont {S.}~\bibnamefont {Kivelson}},
  \bibinfo {author} {\bibfnamefont {M.}~\bibnamefont {Norman}}, \bibinfo
  {author} {\bibfnamefont {S.}~\bibnamefont {Uchida}}, \ and\ \bibinfo {author}
  {\bibfnamefont {J.}~\bibnamefont {Zaanen}},\ }\href {\doibase
  10.1038/nature14165} {\bibfield  {journal} {\bibinfo  {journal} {Nature}\
  }\textbf {\bibinfo {volume} {518}},\ \bibinfo {pages} {179} (\bibinfo {year}
  {2015})}\BibitemShut {NoStop}%
\bibitem [{\citenamefont {Bari{\v{s}}i{\'c}}\ \emph {et~al.}(2015)\citenamefont
  {Bari{\v{s}}i{\'c}}, \citenamefont {Chan}, \citenamefont {Veit},
  \citenamefont {Dorow}, \citenamefont {Ge}, \citenamefont {Tang},
  \citenamefont {Tabis}, \citenamefont {Yu}, \citenamefont {Zhao},\ and\
  \citenamefont {Greven}}]{barivsic2015hidden}%
  \BibitemOpen
  \bibfield  {author} {\bibinfo {author} {\bibfnamefont {N.}~\bibnamefont
  {Bari{\v{s}}i{\'c}}}, \bibinfo {author} {\bibfnamefont {M.}~\bibnamefont
  {Chan}}, \bibinfo {author} {\bibfnamefont {M.}~\bibnamefont {Veit}}, \bibinfo
  {author} {\bibfnamefont {C.}~\bibnamefont {Dorow}}, \bibinfo {author}
  {\bibfnamefont {Y.}~\bibnamefont {Ge}}, \bibinfo {author} {\bibfnamefont
  {Y.}~\bibnamefont {Tang}}, \bibinfo {author} {\bibfnamefont {W.}~\bibnamefont
  {Tabis}}, \bibinfo {author} {\bibfnamefont {G.}~\bibnamefont {Yu}}, \bibinfo
  {author} {\bibfnamefont {X.}~\bibnamefont {Zhao}}, \ and\ \bibinfo {author}
  {\bibfnamefont {M.}~\bibnamefont {Greven}},\ }\href {\doibase
  arxiv.org/abs/1507.07885} {\bibfield  {journal} {\bibinfo  {journal}
  {arXiv:1507.07885}\ } (\bibinfo {year} {2015}),\
  arxiv.org/abs/1507.07885}\BibitemShut {NoStop}%
\bibitem [{\citenamefont {Reyren}\ \emph {et~al.}(2007)\citenamefont {Reyren},
  \citenamefont {Thiel}, \citenamefont {Caviglia}, \citenamefont {Kourkoutis},
  \citenamefont {Hammerl}, \citenamefont {Richter}, \citenamefont {Schneider},
  \citenamefont {Kopp}, \citenamefont {R{\"u}etschi}, \citenamefont {Jaccard},
  \citenamefont {Gabay}, \citenamefont {Muller}, \citenamefont {Triscone},\
  and\ \citenamefont {Mannhart}}]{reyren2007superconducting}%
  \BibitemOpen
  \bibfield  {author} {\bibinfo {author} {\bibfnamefont {N.}~\bibnamefont
  {Reyren}}, \bibinfo {author} {\bibfnamefont {S.}~\bibnamefont {Thiel}},
  \bibinfo {author} {\bibfnamefont {A.~D.}\ \bibnamefont {Caviglia}}, \bibinfo
  {author} {\bibfnamefont {L.~F.}\ \bibnamefont {Kourkoutis}}, \bibinfo
  {author} {\bibfnamefont {G.}~\bibnamefont {Hammerl}}, \bibinfo {author}
  {\bibfnamefont {C.}~\bibnamefont {Richter}}, \bibinfo {author} {\bibfnamefont
  {C.~W.}\ \bibnamefont {Schneider}}, \bibinfo {author} {\bibfnamefont
  {T.}~\bibnamefont {Kopp}}, \bibinfo {author} {\bibfnamefont {A.-S.}\
  \bibnamefont {R{\"u}etschi}}, \bibinfo {author} {\bibfnamefont
  {D.}~\bibnamefont {Jaccard}}, \bibinfo {author} {\bibfnamefont
  {M.}~\bibnamefont {Gabay}}, \bibinfo {author} {\bibfnamefont {D.~A.}\
  \bibnamefont {Muller}}, \bibinfo {author} {\bibfnamefont {J.~M.}\
  \bibnamefont {Triscone}}, \ and\ \bibinfo {author} {\bibfnamefont
  {J.}~\bibnamefont {Mannhart}},\ }\href {\doibase 10.1126/science.1146006}
  {\bibfield  {journal} {\bibinfo  {journal} {Science}\ }\textbf {\bibinfo
  {volume} {317}},\ \bibinfo {pages} {1196} (\bibinfo {year}
  {2007})}\BibitemShut {NoStop}%
\end{thebibliography}
%

\end{document}